\documentclass[12pt]{article}
\usepackage{epsf,amsfonts,amsmath}
\usepackage{graphicx,float,subfigure,epsfig,color,rotating,latexsym}
\usepackage[latin1]{inputenc}
\usepackage[T1]{fontenc}
\def\be{\begin{equation}}
\def\ee{\end{equation}} 
\def\bea{\begin{eqnarray}}
\def\eea{\end{eqnarray}} 
\def\ba{\begin{array}} 
\def\ea{\end{array}}

\begin{document}

\begin{center} 

{\bf \large{Mapping Fermion and Boson systems onto the }}\\
{\bf \large{Fock space of harmonic oscillators}}\\

\vspace*{0.8 cm}

Vincenzo Branchina\footnote{vincenzo.branchina@ct.infn.it}\label{one}
\vspace*{0.4 cm}

Department of Physics, University of
Catania and \\ INFN, Sezione di Catania, 
Via Santa Sofia 64, I-95123, Catania, Italy 

\vspace*{0.6 cm}

Marco Di Liberto\footnote{madiliberto@ssc.unict.it}\label{two}

\vspace*{0.4 cm}

Scuola Superiore di Catania, Via S. Nullo 5/i, Catania, Italy  

\vspace*{0.6 cm}

Ivano Lodato\footnote{ivlodato@ssc.unict.it}\label{three}

\vspace*{0.4 cm}
Scuola Superiore di Catania, Via S. Nullo 5/i, Catania, Italy and\\
INFN, Sezione di Catania, Via Santa Sofia 64, I-95123, Catania, Italy 

\vspace*{1.2 cm}

{\LARGE Abstract}\\

\end{center} 
The fluctuation-dissipation theorem (FDT) is very general and applies 
to a broad variety of different physical phenomena in  condensed matter 
physics. With the help of the FDT and following the famous work of 
Caldeira and Leggett, we show that, whenever linear response theory 
applies, any generic bosonic or fermionic system at finite temperature 
$T$ can be mapped onto a fictitious system of free harmonic oscillators. 
To the best of our knowledge, this is the first time that such a mapping 
is explicitly worked out. This finding provides further theoretical 
support to the phenomenological harmonic oscillator models commonly used 
in condensed matter. Moreover, our result helps in 
clarifying an interpretation issue related to the presence and physical 
origin of the Bose-Einstein factor in the FDT.  

\section{Introduction}

The idea of modeling physical systems as a collection of harmonic 
oscillators has a long history and dates back to even before the 
birth of quantum mechanics. One of the best known example is  
Planck's work on black body radiation at the edge 
of the classical (beginning  of quantum) era. More recently, this 
model has been of great importance in connection with the study 
of dissipation in quantum mechanics\,\cite{seni, ford, mohri}.

In their famous paper devoted to the study of tunneling in 
dissipative systems\,\cite{caleg2}, Caldeira and Legget observed that 
``any physical system which is weakly perturbed around its equilibrium 
state can be adequately represented (at T=0 at least) by regarding that 
system as equivalent to a set of simple harmonic oscillators''. 
They supplemented this statement with an explicit computation, where they 
showed that given a quantum system at $T=0$, in the lowest order 
approximation, its dynamics can be reproduced with the help 
of a properly constructed system of harmonic oscillators. They also
stressed that the study of the $T \neq 0$ case (not considered
in their paper) needed separate discussion. 

Caldeira and Legget then suggested the form of 
the total lagrangian of a physical system in interaction with a certain 
``environment'' (the ``heat bath'') and worked out the consequences of 
this assumption in connection with the tunneling 
problem\,\cite{caleg2, caleg1}.  
Prompted by this pioneering work, harmonic oscillator models are nowadays
extensively used and there is little doubt that, from a phenomenological 
point of view, they reproduce quite well the physics of the systems 
under investigation. As a specific example, we can consider an electrical 
circuit, where the resistance is modeled with a collection of 
harmonically oscillating electrical dipoles.  

From a theoretical point of view, however, it would be more satisfactory 
if we could prove that, for any generic system at finite temperature 
$T \neq 0$, it is possible to find an equivalent system of harmonic 
oscillators such that the statistical (thermodynamical) properties of 
the real physical system are properly reproduced by the system of 
oscillators. This would be an extension of the Caldeira Legget 
result and would provide further theoretical support to the commonly 
used phenomenological models. 

The main scope of this work is to present a new and very 
general result which provides the above mentioned extension of the 
Caldeira-Legget one. By working within the framework of the 
FDT\,\cite{cawe}, we show that, whenever linear response theory applies, 
any generic bosonic and/or fermionic system at finite temperature can be 
mapped onto the Fock space of a fictitious system of free harmonic 
oscillators at the same temperature.

As a byproduct of our analysis, we shall see that our finding 
should help in clarifying an interpretation issue concerning the Bose-Einstein 
(BE) distribution factor which appears in the FDT. Actually, an often 
raised question concerns the physical meaning and/or origin of the BE 
factor which appears in the relation between the power 
spectrum of the fluctuating quantity and the corresponding generalized 
susceptibility. 
Sometimes this term is interpreted as due to an harmonic oscillator 
composition of the physical system under investigation. Such an 
interpretation, however, is not supported by the derivation of the 
theorem itself (see for instance\,\cite{kubo, jetz2, noi}). 
Moreover, the FDT applies to any generic bosonic or fermionic
system (irrespectively of its statistics). 

Far from being an academic question, the resolution of this 
interpretation issue is of very practical importance in many different 
contexts\,\cite{koch, koch1, bema1, bema2}. From a 
real understanding of the origin of this term often depends the correct 
physical interpretation of theoretical and experimental 
results\,\cite{jetz2, noi, jetz1, doran, maha, taylor}. As we shall 
see, our results suggest that this term does not originate from  
underlying physical oscillator degrees of freedom of the system but 
is rather a general property related to the approximation (linear 
response) involved in the derivation of the FDT. 

According to this theorem, whenever linear response theory is applicable,
given a generic system which interacts with an external field $f(t)$ 
through the interaction term $\hat V = - f(t)\,\hat A$, where 
$\hat A$ is an observable of the system, the mean square of the Fourier 
transform $\hat A(\omega)$ of $\hat A(t)$ is related to the imaginary 
part $\chi_{_A}''(\omega)$ of the corresponding (Fourier transformed) 
generalized susceptibility by the relation\,\cite{cawe} :
\be \label{fdt1} 
\langle \hat{A}^2(\omega)\rangle =
\hbar \chi_{_A}^{\,''}(\omega) \,{\rm coth}\left( \frac{\beta\hbar\omega}{2}
\right) =2\,\hbar\, \chi_{_A}^{\,''}(\omega) \,\left(\frac1 2 + \frac{1}
{e^{\beta\hbar\omega}-1}\right)\,,
\ee
where $\beta=1/k T$, with $T$ the temperature of the system and $k$ the 
Boltzmann constant.

For instance, in the case of a resistively shunted Josephson 
junction\,\cite{koch}, when applied to the power spectrum 
$S_I(\omega)$ of the noise current (fluctuation) in the resistive 
shunt (dissipation), the theorem takes the form 
($R$ is the shunt resistance)\cite{koch1}:
\be
\label{spectr}
S_I(\omega) = \frac{4}{R}\left(\frac{\hbar\omega}{2}+
\frac{\hbar\omega}{e^{\beta\hbar\omega}-1}\right)\, .
\ee
The power spectrum $S_I(\omega)$ has been measured\,\cite{koch}
and good agreement between the experimental results and 
Eq.\,(\ref{spectr}) has been found. 

The above $\frac{\hbar\omega}{2}$ term is sometimes 
presented\,\cite{koch, koch1, gardizo, kogan} as due to zero point 
energies and the experimental results\,\cite{koch} as a measurement 
of them. In fact, the term in parenthesis in 
Eq.\,(\ref{spectr}) coincides with the mean energy 
of an harmonic oscillator of frequency $\omega$ in a thermal bath. 
The same holds true for the general case of Eq.\,(\ref{fdt1}),
where the similar term is the mean energy of an harmonic 
oscillator in $\hbar\omega$ units, i.e.\,the BE distribution function. 
In the following we shall see that our result (the mapping) strongly 
suggests that the agreement between the experimental results\,\cite{koch} 
and Eq.\,(\ref{spectr}) cannot be considered as a signature of measurement 
of zero point energies. 

The rest of the paper is organized as follows. In section 2 we 
briefly review the derivation of the FDT and establish some relations 
useful for the following. In section 3 we establish our new result, the
{\it mapping}, i.e. we show that for any given bosonic or fermionic system 
at finite temperature $T \neq 0$ we 
can always find a fictitious system of harmonic oscillators in such a 
manner that the physical quantities which appear in the FDT can be 
obtained from this equivalent system of oscillators. Section 4 is for 
our comments and conclusions. In particular, in this last section we 
present our comments on the interpretation issue related to the BE 
term in the FDT.  

\section{The fluctuation-dissipation theorem}

Let us begin by briefly reviewing the derivation 
of the FDT. Consider a macroscopic system with unperturbed 
Hamiltonian $\hat{H}_0$ under the influence of the perturbation 
\be
\label{inter}
\hat{V} = - f(t)\,\hat {A}(t)\,,
\ee
where $\hat{A}(t)$ is an observable (a bosonic operator) of 
the system and $f(t)$ an external 
generalized force\footnote{More generally, we could consider 
a local observable and a local generalized force, in which case 
we would have $\hat{V} = -\int d^3\,\vec r \hat{A}(\vec{r})f(\vec{r},t)$, 
and successively define a local susceptibility 
$\chi(\vec{r},t;\vec{r'},t')$ (see Eq.\,(\ref{chi}) below).  
As this would add nothing to our argument, we shall restrict ourselves
to $\vec r$-independent quantities. The extension 
to include local operators is immediate.}. 
Let $|E_n\rangle$ be the $\hat{H}_0$ 
eigenstates (with eigenvalues $E_n$) and  
$\langle E_n|\hat{A}(t)|E_n \rangle =0$. 
Within the framework of linear response theory, the quantum-statistical 
average $\langle\hat{A}(t)\rangle_f$ of the observable $\hat{A}(t)$ 
in the presence of $\hat{V}$ is given by 
\be
\label{resp2}
\langle \hat{A}(t)\rangle_f = \int_{-\infty}^t\, d t' \chi_{_{A}}(t-t') f(t') 
\ee
where $\chi_{_{A}}(t - t')$ is the generalized susceptibility,
\be
\label{chi}
\chi_{_{A}}(t - t')=\frac{i}{\hbar}\theta(t-t') 
\langle [\hat{A}(t),\hat{A}(t')] \rangle = 
-\frac{1}{\hbar}G_R(t - t')\, ,
\ee
with $\langle ... \rangle = 
\sum_{n} \varrho_n \langle E_n| ... |E_n \rangle$,\, 
$\varrho_n= e^{-\beta E_n}/Z$\, ,  $Z=\sum_n e^{- \beta E_n}$\,, 
$G_R(t-t')$ being the retarded Green's function and 
$\hat{A}(t)=e^{i\hat{H_0}t/\hbar}\hat{A}e^{-i\hat{H_0}t/\hbar}$.

Defining the correlators (from now on  $t^{'}=0$):
\be\label{correla}
G_{>}(t)= \langle \hat{A}(t)\,\hat{A}(0) \rangle 
\,\,\,\,\,\,\,\,\,\,\,\,\,\,\, {\rm and} \,\,\,\,\,\,\,\,\,\,\,\,\,\,\, 
G_{<}(t)= \langle \hat{A}(0)\,\hat{A}(t) \rangle\,, 
\ee
so that $ G_R(t)=-i\theta(t)(G_{>}(t) - G_{<}(t))$, 
and the corresponding Fourier transforms,
$ G_{>}(\omega)$ and $ G_{<}(\omega)$ respectively, it is a matter of 
few lines to show that:
\be\label{aux}
 G_{>}(\omega)=-\frac{2}{1-e^{-\beta\hbar\omega}}{\rm Im}\,G_R(\omega)
\,\,\,\,\,\,\,\,\,\,\,\,\,\,\, ; \,\,\,\,\,\,\,\,\,\,\,\,\,\,\, 
G_{<}(\omega)=e^{-\beta\hbar\omega}\,G_{>}(\omega) \,.
\ee
Finally, by noting that  
\be\label {oo2}
\langle \hat{A}^2(\omega)\rangle= \frac12 (G_{>}(\omega) + G_{<}(\omega)) 
\ee
and that the Fourier transform of $\chi_{_{A}}(t)$\, is\, 
$\chi_{_{A}}(\omega) = \chi_{_{A}}^{\,'}(\omega) + i \chi_{_{A}}^{\,''}(\omega) = 
-\frac{1}{\hbar} G_{R}(\omega)$ we get:
\be \label{fddt} 
\langle \hat{A}^2(\omega)\rangle = \hbar \chi_{_{A}}^{\,''}(\omega) \, \frac
{1 + e^{- \beta\hbar\omega}}
{1 - e^{- \beta\hbar\omega}} =
\hbar \chi_{_{A}}^{\,''}(\omega) \,{\rm coth}\left( \frac{\beta\hbar\omega}{2}\right)
=2\,\hbar \chi_{_{A}}^{\,''}(\omega) \,
\left(\frac1 2 + \frac{1}{e^{\beta\hbar\omega}-1}\right) 
\ee
which is Eq.\,(\ref{fdt1}), the celebrated FDT. 

As observed by Kubo et al.\,\cite{kubo} (and shown in the derivation sketched 
above), the BE factor is simply due to a peculiar combination of Boltzmann 
factors in Eq.\,(\ref{fddt}) and there is no reference to physical harmonic 
oscillators of the system whatsoever.
Despite such an authoritative remark, some people insist in interpreting 
the $\left(\frac1 2 + \frac{1}{e^{\beta\hbar\omega}-1}\right)$ term as related 
to harmonic oscillator degrees of freedom of the physical system.

In the case of the measured\,\cite{koch} power spectrum of Eq.\,(\ref{spectr}), 
some authors\,\cite{bema1, bema2} interpret this term as due to the 
electromagnetic field in the resistive shunt and therefore the first term 
in parenthesis of Eq.\,(\ref{spectr}) as originating from zero point energies 
of this electromagnetic field.
Such an interpretation, however, is not supported by any physical derivation
and has been strongly criticized in\,\cite{jetz2, jetz1, doran, maha}.
Very recently, starting from the results of the present work, we have also 
carefully investigated this issue\,\cite{noi}, providing arguments which 
strongly support previous criticisms (see\,\cite{noi} for details).

Let us go back now to our analysis. For our purposes,  
it is useful to show that from Eqs.\,(\ref{chi}) 
and (\ref{oo2}) we can easily derive the following expressions for 
$\chi_{_A}^{\,''}(\omega)$ and $\langle \hat{A}^2(\omega)\rangle$\,:      
\be\label{chi2}
\chi_{_A}''(\omega)=\frac{\pi}{\hbar}\sum_{i, j}\varrho_i |A_{i j}|^2
\left[\delta\left(\frac{E_i-E_j}{\hbar}+\omega\right)-
\delta\left(\frac{E_j-E_i}{\hbar}+\omega\right)\right]\,,
\ee
and 
\be\label{o2}
\langle \hat{A}^2(\omega)\rangle=\pi\sum_{i, j}\varrho_i |A_{ij}|^2
\left[\delta\left(\frac{E_i-E_j}{\hbar}+\omega\right)+
\delta\left(\frac{E_j-E_i}{\hbar}+\omega\right)\right]\,,
\ee
where $A_{ij}=\langle E_i|\hat A | E_j\rangle $ . 

In fact, by inserting in Eq.\,(\ref{chi}) the expressions:\hfill\break
$\theta(t-t')=-\int_{-\infty}^{+\infty}\frac{d\omega}
{2\pi i}\frac{e^{-i\omega(t-t')}}{\omega+i\eta}$ , $I=\sum_i |E_i\rangle\langle E_i|$
and $\hat{A}(t)=e^{i\hat{H_0}t/\hbar}\hat{A}e^{-i\hat{H_0}t/\hbar}$ we get:
\begin{equation}
\chi_{_A}(t-t')=-\frac1 \hbar \int \frac{d\omega}{2\pi}\frac{e^{-i\omega(t-t')}}{\omega+i\eta}
\sum_{i, j} \varrho_i |A_{ij}|^2
\left(e^{i(E_i-E_j)(t-t')/\hbar}- e^{-i(E_i-E_j)(t-t')/\hbar} \right)
\end{equation}
Then, making use of 
$\lim_{\eta\rightarrow 0}\frac{1}{\omega+i\eta}= \mathcal{P} 
\left(\frac1 \omega\right)  - i\pi\delta(\omega)$\,, Eq.\,(\ref{chi2}) follows immediately. 

As for Eq.\,(\ref{o2}), from Eq.\,(\ref{correla}) for $G_>(t)$, we have: 
\be\label{uno}
G_>(t) = \sum_{i,j}\rho_i
\langle E_i| e^{\frac i \hbar \hat H t}\hat A 
e^{-\frac i \hbar \hat H t}|E_j\rangle\langle E_j|\hat A|E_i\rangle=\nonumber \\ 
=\sum_{i,j}\rho_i e^{-\frac i \hbar(E_j - E_i)t}|A_{ij}|^2\,.
\ee
Working out the similar expression for $G_<(t)$, for the correlation function 
$G(t)$ we get: 

\begin{equation}\label{due}
G(t)=\frac1 2 (G_>(t) + G_<(t))=\frac1 2 \sum_{i , j}\rho_i |A_{ij}|^2 
(e^{-\frac i \hbar(E_j - E_i)t}+e^{-\frac i \hbar(E_i - E_j)t})\, ,
\end{equation}
so that the Fourier transform $\tilde G(\omega)$ is:
\be\label{ft}
\tilde G(\omega) =\pi\sum_{i , j}\rho_i|A_{ij}|^2\left[\delta\left(\frac{E_j-E_i}{\hbar}+
\omega\right)+\delta\left(\frac{E_i-E_j}{\hbar} +\omega\right) \right]\,.
\ee
As $\langle\hat{A}^2\rangle = G(0)$, $\tilde G(\omega)$ is the spectral 
density $\langle \hat{A}^2(\omega)\rangle$ of $\langle\hat{A}^2\rangle$\, 
(\,$\langle \hat{A}^2\rangle=\int_{-\infty}^{+\infty}\langle \hat{A}^2(\omega)
\rangle\frac{d\omega}{2\pi}$\,). Then, making use of Eq.\,(\ref{ft}), we 
finally get Eq.\,(\ref{o2}). 

For our purposes, it is also useful to write Eq.\,(\ref{o2}) in
a different manner. After some straightforward manipulations, 
Eq.\,(\ref{o2}) can be written as:
\bea
\langle \hat{A}^2(\omega)\rangle &=& \pi\sum_{j > i}(\varrho_i - \varrho_j) |A_{ij}|^2
\,{\rm coth}\left( \frac{\beta\hbar\omega_{ji}}{2}\right)
\left[\delta\left(\omega -\omega_{ji}\right)+
\delta\left(\omega + \omega_{ji}\right)\right]\label{o5}\\
&=& \pi\,{\rm coth}\left(\frac{\beta\hbar\omega}{2}\right)\sum_{j > i}
(\varrho_i - \varrho_j) |A_{ij}|^2 \left[\delta\left(\omega -\omega_{ji}\right)-
\delta\left(\omega + \omega_{ji}\right)\right]\,,\label{o6}
\eea
where we have introduced the notation: $\omega_{ji}=\frac{E_j-E_i}{\hbar}$.
By following similar steps, Eq.\,(\ref{chi2}) can also be written as:  
\be\label{chii3}
\chi''(\omega)=\frac{\pi}{\hbar}\sum_{j > i}(\varrho_i - \varrho_j) |A_{i j}|^2
\left[\delta\left(\omega - \omega_{ji}\right)-
\delta\left(\omega + \omega_{ji}\right)\right]\,. 
\ee
Clearly, comparing Eq.\,(\ref{o6}) with Eq.\,(\ref{chii3}), we find, as 
we should, the FDT theorem. 

Now, starting from Eqs.\,(\ref{o5}) and (\ref{chii3}) and taking inspiration 
from the seminal work of Caldeira and Leggett\,\cite{caleg2}, we shall 
be able to establish a formal mapping between the real system considered 
so far and a system of fictitious harmonic oscillators. 
A similar mapping, restricted however to the $T=0$ case, 
was considered in\,\,\cite{caleg2}, where it was also noted that 
the $T\neq 0$ case needs separate discussion. 
The mapping that we are going to construct in the present work  
deals with the $T\neq 0$ general case. 

\section{Mapping Boson and Fermion systems onto harmonic oscillators}

To prepare the basis for the construction of this mapping, let us 
consider first a real system ${\cal S}_{osc}$ of 
harmonic oscillators (each of which is labeled below by the double 
index $\{ji\}$ for reasons that will become clear in the following) 
whose free Hamiltonian is: 
\be\label{armonico}
\hat H_{osc} = \sum_{j > i}\left(\frac{\hat p_{ji}^{\,2} }{2 M_{ji}} + 
\frac{M_{ji} \omega_{ji}^2}{2}\,\hat q_{ji}^{\,2}\right)\,,
\ee
where $\omega_{ji}$ are the proper 
frequencies of the individual harmonic oscillators and $M_{ji}$ their 
masses. Let  $| n_{j i}\rangle$ ($n_{ji}=0, 1,2,...$) be the 
occupation number states of the $\{ji\}$ oscillator out of which the Fock 
space of ${\cal S}_{osc}$ is built up.
Let us consider also ${\cal S}_{osc}$ in interaction with an external 
system through the one-particle operator: 
\be\label{armint}
\hat V_{osc} = - f(t) \hat A_{osc}\,,
\ee
with 
\be\label{onepart}
{\hat A}_{osc} = \sum_{j > i} \left(\alpha_{j i} \,{\hat q}_{ji} \right)\,.
\ee
Obviously, the FDT applied to ${\cal S}_{osc}$ gives 
$\langle {\hat A}_{osc}^2(\omega)\rangle = 
\hbar \chi_{osc}^{\,''}(\omega) \,{\rm coth}\left( \frac{\beta\hbar\omega}{2}\right)$\,,
but this is not what matters to us. 

What is important for our purposes is that, differently from 
any other generic system, for ${\cal S}_{osc}$ we can exactly compute 
$\langle {\hat A}_{osc}^2(\omega) \rangle$ and $\chi_{osc}^{\,''}(\omega)$
from Eqs.\,(\ref {chi2}) and (\ref{o2}) because we can explicitly compute 
the matrix elements of ${\hat A}_{osc}$. 

In fact, if we apply Eqs.\,(\ref{uno}) and (\ref{ft}) 
to ${\cal S}_{osc}$ and replace the double index 
notation 
($\omega_{ji}$ ; $n_{ji}$ ; $M_{ji}$ ;  etc.) with the more 
convenient (for the time being) and self explanatory one index 
notation ($\omega_{1}$, $\omega_{2}$, ... ; 
$n_{1}$, $n_{2}$, ... ; $M_{1}$, $M_{2}$, ... ; etc.), for 
$\langle \hat{A}_{osc}^2(\omega)\rangle$ we have:
\bea\label{osc}
\langle \hat{A}_{osc}^2(\omega)\rangle&=&\pi
\sum_{n_1,n_2,..}\,\,\,\,  \sum_{m_1,m_2,..}\,(\varrho_{n_1}  \varrho_{n_2}\cdots)
|\langle n_1,n_2,..|\hat{A}_{osc} | m_1,m_2,..\rangle|^2 \nonumber\\
&&\times
\left[\delta \left(\omega + l_1\omega_1 + l_2\omega_2 +\cdots \right)+ 
\delta \left(\omega - l_1\omega_1 - l_2\omega_2 - \cdots \right)\right]\,,
\eea
where $l_k=n_k-m_k$, $\varrho_{n_k} = e^{-\beta (n_k + 1/2)\hbar\omega_k}/Z_k$, 
$Z_k = \sum_{n_k} e^{-\beta (n_k + 1/2)\hbar\omega_k}$ (note also that in this 
one index notation  $\hat{A}_{osc}$ is written as \, 
${\hat A}_{osc} = \sum_{k} \left(\alpha_{k} \,{\hat q}_{k} \right)$\,). 
Now, as
\be
\langle n_k| {\hat q}_{k} | m_k\rangle = \sqrt {\frac{\hbar}{2 M_k \omega_k}}
\left(\sqrt {n_k + 1} \langle n_k + 1| m_k \rangle + 
\sqrt {n_k} \langle n_k - 1| m_k \rangle \right) \,, 
\ee
we immediately get:
\bea\label{osc2}
\langle \hat{A}_{osc}^2(\omega)\rangle&=&\pi
\sum_{n_1,n_2,..}\,\,\,\,  \sum_{m_1,m_2,..}\,(\varrho_{n_1}  \varrho_{n_2}\cdots)
~~~~~~~~~~~~~~~~~~~~~~~~~~~~~~~~~~~~~~~~~~~~~~~~~~~~~~~~~~~~\nonumber\\
&\times&\left[\sum_k \alpha_k \sqrt {\frac{\hbar}{2 M_k \omega_k}}
\left(\sqrt {n_k + 1} \delta_{m_k, n_k+1} + \sqrt {n_k} \delta_{m_k, n_k-1}\right)
\prod_{h\neq k} \delta_{m_h, n_h}\right]^2 \nonumber\\
&\times&
\left[\delta \left(\omega + l_1\omega_1 + l_2\omega_2 +\cdots \right)+ 
\delta \left(\omega - l_1\omega_1 - l_2\omega_2 - \cdots \right) \right]\,.
\eea

Let us concentrate our attention to the square in 
the second line of Eq.\,(\ref{osc2}). Due to the presence of the 
Kronecker deltas,  
all the crossed terms in this square, i.e. all the terms with different 
values of the index $k$, vanish. In other words, the square of the sum 
is equal to the sum  of the squares:
\bea
\left[\sum_k \alpha_k \sqrt {\frac{\hbar}{2 M_k \omega_k}}
\left(\sqrt {n_k + 1} \delta_{m_k, n_k+1} + \sqrt {n_k} \delta_{m_k, n_k-1}\right)
\prod_{h\neq k} \delta_{m_h, n_h}\right]^2 \nonumber\\
= \sum_k \left(\alpha_k \sqrt {\frac{\hbar}{2 M_k \omega_k}}
\left(\sqrt {n_k + 1} \delta_{m_k, n_k+1} + \sqrt {n_k} \delta_{m_k, n_k-1}\right)
\prod_{h\neq k} \delta_{m_h, n_h}\right)^2
\eea
For the same reason, the same holds true
for each value of the index $k$, i.e.:
\bea
&&\left(\alpha_k \sqrt {\frac{\hbar}{2 M_k \omega_k}}
\left(\sqrt {n_k + 1} \delta_{m_k, n_k+1} + \sqrt {n_k} \delta_{m_k, n_k-1}\right)
\prod_{h\neq k} \delta_{m_h, n_h}\right)^2 \nonumber\\
&=&\alpha_k^2\, \frac{\hbar}{2 M_k \omega_k}
\left( (n_k + 1)\, \delta_{m_k, n_k+1} + n_k\, \delta_{m_k, n_k-1}\right)
\prod_{h\neq k} \delta_{m_h, n_h}\label{osscc}\,.
\eea
Therefore, as $l_k=n_k-m_k $, for 
$\langle \hat{A}_{osc}^2(\omega)\rangle$ we get:
\be\label{osc3}
\langle \hat{A}_{osc}^2(\omega)\rangle=\pi
\sum_{n_1,n_2,..}\, (\varrho_{n_1}  \varrho_{n_2}\cdots)\,
\sum_k \alpha_k^2\, \frac{\hbar}{2 M_k \omega_k}
(2\,n_k + 1)\, (\delta (\omega-\omega_k) +\delta (\omega+\omega_k) )\,.
\ee
Finally, as $ \sum_{n_k}\, \varrho_{n_k}=1 $, the above expression becomes: 
\bea
\langle \hat{A}_{osc}^2(\omega)\rangle&=&\pi
\sum_k \, \alpha_k^2\, \frac{\hbar}{2 M_k \omega_k}
(\delta (\omega-\omega_k) +\delta (\omega+\omega_k))
\sum_{n_k}\, \varrho_{n_k}\,(2\,n_k + 1) \label{si}\\
&=&\pi
\sum_k \, \alpha_k^2\, \frac{\hbar}{2 M_k \omega_k}\,
{\rm coth}\left(\frac{\beta \hbar\omega_k}{2}\right)
(\delta (\omega-\omega_k) +\delta (\omega+\omega_k))\label{osc4}\,.
\eea
Going back to the original double index notation: 
\bea
\langle \hat{A}_{osc}^2(\omega)\rangle&=&\pi
\sum_{j>i} \, \alpha_{ji}^2\, \frac{\hbar}{2 M_{ji} \omega_{ji}}\,
{\rm coth}\left(\frac{\beta \hbar\omega_{ji}}{2}\right)
(\delta (\omega-\omega_{ji}) +\delta (\omega+\omega_{ji}))\label{osc5}\\
&=& \pi\, {\rm coth}\left(\frac{\beta \hbar\omega}{2}\right)\sum_{j>i} \,
\alpha_{ji}^2\,\frac{\hbar}{2 M_{ji} \omega_{ji}}\,
(\delta (\omega-\omega_{ji}) - \delta (\omega+\omega_{ji}))\label{oscc5}\,.
\eea

We have just seen that given a real system ${\cal S}_{osc}$ of harmonic 
oscillators and the one particle operator $\hat{A}_{osc}$  of 
Eq.\,(\ref{onepart}), for such an operator is possible to evaluate explicitly 
$\langle \hat{A}_{osc}^2(\omega)\rangle$. We find that each of the individual 
harmonic oscillators gives rise to a term  
${\rm coth}\left(\frac{\beta \hbar\omega_{ji}}{2}\right)$ which in turn 
comes from the term $\sum_{n_{ji}}\, \varrho_{n_{ji}}\,(2\,n_{ji} + 1)$
of Eq.\,(\ref{si}).

Let us now consider $\chi_{osc}^{''} (\omega)$,  which   
(see Eqs.\,(\ref{chi2}) and (\ref{osc2})) is nothing but:
\bea\label{cchi2}
\chi^{''}_{osc} (\omega)&=&\frac{\pi}{\hbar}
\sum_{n_1,n_2,..}\,\,\,\,  \sum_{m_1,m_2,..}\,(\varrho_{n_1}  \varrho_{n_2}\cdots)
~~~~~~~~~~~~~~~~~~~~~~~~~~~~~~~~~~~~~~~~~~~~~~~~~~~~~~~~~~~~\nonumber\\
&\times&\left[\sum_k \alpha_k \sqrt {\frac{\hbar}{2 M_k \omega_k}}
\left(\sqrt {n_k + 1} \delta_{m_k, n_k+1} + \sqrt {n_k} \delta_{m_k, n_k-1}\right)
\prod_{h\neq k} \delta_{m_h, n_h}\right]^2 \nonumber\\
&\times&
\left[\delta \left(\omega + l_1\omega_1 + l_2\omega_2 +\cdots \right) -  
\delta \left(\omega - l_1\omega_1 - l_2\omega_2 - \cdots \right) \right] .
\eea
Apart from the factor $1/\hbar$, Eq.\,(\ref{cchi2}) differs from Eq.\,(\ref{osc2})
because it contains the difference (rather than the sum) of delta functions 
in the last line.  

If we proceed for $\chi_{osc}^{''} (\omega)$ as we have just done for 
$\langle \hat{A}_{osc}^2(\omega)\rangle$, we immediately note that 
the only difference with the previous computation is due to this minus 
sign. In fact, its presence causes that rather than the 
combination  $(2\,n_k + 1)$ of Eq.\,(\ref{osc3}), which comes from the sum 
$(n_k + 1) + n_k$ of Eq.\,(\ref{osscc}), we get the combination
$(n_k + 1) - n_k = 1$. Therefore, for $\chi_{osc}^{''} (\omega)$ we
do not get the sum 
$\sum_{n_k}\, \varrho_{n_k}\,(2\,n_k + 1)= 
{\rm coth}\left(\frac{\beta \hbar\omega_k}{2}\right)$ of Eq.\,(\ref{si}), 
but rather\, $\sum_{n_k}\, \varrho_{n_k}\, = 1$. Then: 
\be\label{chi3}
\chi_{osc}^{''} (\omega)=\frac{\pi}{\hbar}
\sum_{j>i} \,\alpha_{ji}^2\,\frac{\hbar}{2 M_{ji} \omega_{ji}}\,
(\delta (\omega-\omega_{ji}) - \delta (\omega+\omega_{ji}))\,.
\ee

Naturally, comparing Eq.\,(\ref{oscc5}) with Eq.\,(\ref{chi3}) we 
see that for ${\cal S}_{osc}$ the FDT holds true, 
as it should. However, what is important for our purposes is to note 
that for this system we have been able to compute separately 
$\langle \hat{A}_{osc}^2(\omega)\rangle$ and $\chi_{osc}^{''} (\omega)$  
and found that the 
${\rm coth}\left(\frac{\beta \hbar\omega}{2}\right)$ factor of the FDT 
originates from the individual contributions 
${\rm coth}\left(\frac{\beta \hbar\omega_{ji}}{2}\right)$ of each of 
the harmonic oscillators of ${\cal S}_{osc}$.

We are now in the position to build up our mapping. 
Let us consider the original system ${\cal S}$, described by the 
unperturbed Hamiltonian $\hat H_0$\,, in interaction with an external 
field $f(t)$ through the interaction term  $\hat V = - f (t)\,\hat A$  
(see Eq.\,(\ref{inter})), and construct a fictitious system of harmonic 
oscillators ${\cal S}_{osc}$, described by the free Hamiltonian 
${\hat H}_{osc}$ of Eq.\,(\ref{armonico}), in interaction with the same 
external field $f(t)$ through the interaction term ${\hat V}_{osc}$ of 
Eq.\,(\ref{armint}), with $\hat A_{osc}$ given by Eq.\,(\ref{onepart}),  
where for $\alpha_{j i}$ we choose:
\be\label{alfa}
\alpha_{j i} = \left(\frac{2 M_{ji} \omega_{ji}}{\hbar}\right)^{\frac12}
(\varrho_i - \varrho_j)^{\frac12} \,|A_{ij}|
\ee
and for the proper frequencies $\omega_{ji}$ of the oscillators: 
\be\label{omega}
\omega_{ji}= (E_j-E_i)/\hbar > 0\,,
\ee
with $E_i$ the eigenvalues of the Hamiltonian ${\hat H}_0$ of the real system. 

Comparing Eq.\,(\ref{oscc5}) with Eq.\,(\ref{o6}) and Eq.\,(\ref{chi3}) 
with Eq.\,(\ref{chii3}), it is immediate to see that 
with the above choices of $\alpha_{j i}$ and $\omega_{ji}$ we have: 
\be\label{cen1}
\langle \hat{A}^2(\omega)\rangle = \langle \hat{A}_{osc}^2(\omega)\rangle
\,\,\,\,\,\,\,\,\,\,\, {\rm and} \,\,\,\,\,\,\,\,\,\,\,
\chi_{_A}^{''} (\omega) = \chi_{osc}^{''} (\omega)\,.
\ee

Eqs.\,(\ref{alfa}) and (\ref{omega}) are the central results of our 
analysis. Actually, these are the equations which allow to establish our
mapping. In fact, with such a choice of the $\alpha$'s and the $\omega$'s, 
we are able to map the real system ${\cal S}$ onto a fictitious system of 
harmonic oscillators ${\cal S}_{osc}$\,, 
\be
{\cal S} \,\, \to \,\, {\cal S}_{osc}\,,
\ee
in such a manner that $\chi_{_A}^{''} (\omega)$ and 
$\langle \hat{A}^2(\omega)\rangle$ of the real system are equivalently 
obtained by computing the corresponding quantities of the 
fictitious one (Eqs.\,(\ref{cen1})).

This is the desired result. What we have just shown is that any generic 
boson or fermion system at finite temperature $T$ is equivalent to a 
system of harmonic oscillators at the same temperature. 
From a theoretical point of view, the relevance of 
such a result should be immediately clear. As we have already observed, 
in fact, harmonic oscillator models are quite common in modeling 
generic physical systems. Now, the typical physical situation we have to 
deal with is that of a system (bosonic or fermionic) at finite 
temperature $T \neq 0$. 
In this respect, our mapping fills up the gap mentioned by Caldeira and 
Legget (see Appendix C of\,\cite{caleg2}) by extending the $T=0$ mapping 
put forward by them to the general finite temperature case, thus
providing further theoretical support to the use of these models.

In the following section, we would like to add some more comments on the 
above results. In particular, we are going to consider the previously 
mentioned interpretation issue concerning the presence and origin of the 
BE distribution factor in the FDT.

\section{Comments and conclusions} 

First of all, we point out that, in order to 
construct the above mapping, the key ingredient we used is the hypothesis 
that linear response theory is applicable, which is the main hypothesis 
under which the FDT is established. When this is not the case, 
Eq.(\ref{resp2}) cannot be derived and we do not arrive to 
Eqs.\,(\ref{chi2}) and (\ref{o2}), which are crucial to build up our 
mapping.

Note now that, by considering the ``equivalent'' harmonic oscillators system  
${\cal S}_{osc}$ rather than the real one, we are somehow allowed to 
regard the BE distribution factor 
${\rm coth}\left(\frac{\beta\hbar\omega}{2}\right)$ 
of the FDT in Eq.\,(\ref{fdt1}) as originating from the individual
contributions ${\rm coth}\left(\frac{\beta\hbar\omega_{ji}}{2}\right)$ 
of each of the oscillators of the fictitious system (see above,
Eqs.\,(\ref{osc5}), (\ref{oscc5}) and (\ref{chi3})). 
In this sense, such a mapping allows for an oscillator 
interpretation of the BE term in the FDT.

At the same time, however, our result shows that this BE factor does 
not describe the physics of the system, i.e.\,\,it does not encode any 
real, physical, harmonic oscillator degrees of freedom of the system 
(see also the considerations below). 

In this respect, it is worth to point out that what we have implemented  
is not a canonical transformation, i.e.\,it is not a transformation which 
allows to describe the system in terms of new degrees of freedom 
(such as normal modes), 
but a formal mapping, a mathematical construct, which can be 
established, we repeat ourselves, only within the framework of linear 
response theory.

In our opinion, then, our finding provides an answer to the 
questions of the ``physical meaning'' or ``physical origin'' 
of the BE term in the FDT or, stated differently, to the question of 
whether this BE distribution factor possibly describes the physical 
nature of the system or not\,\cite{taylor}. 

In fact, from the derivation of the FDT, we know 
that the BE factor derives from a peculiar combination of Boltzmann 
factors (see\,\cite{kubo} and Eq.\,(\ref{fddt}) above). At the same 
time, we have shown that, regardless the bosonic or fermionic 
nature of the (real) system ${\cal S}$, it is always possible to
establish a mapping which relates ${\cal S}$ to a system of harmonic 
oscillators ${\cal S}_{osc}$ so that this BE factor can be regarded 
as ``originating'' from the individual oscillators of the 
``equivalent'' system ${\cal S}_{osc}$. 
Therefore, it is not the physical nature 
of the system which is encoded in this BE term but rather a fundamental 
quantum property of any bosonic and/or fermionic system: 
{\it whenever 
linear response theory is applicable, any generic system is, at least 
with respect to the FDT, equivalent (in the sense defined above) to 
a system of quantum harmonic oscillators}.

Before ending this section, it is probably worth to spend few words
on some examples of realistic systems where our mapping is at work. 
In this respect, we would like to note that there are several applications 
in the literature where fermionic systems, after bosonization, are actually 
described by a system in interaction with a bosonic bath. This is, for 
instance, the case of the anisotropic Kondo model, which is shown to be 
equivalent to a spin-boson model (a two level system in interaction with 
a bosonic bath).
The same is also true for a quantum dot interacting with external leads. 

In the above examples, the system is described with the 
help of a spin-boson Hamiltonian, thus providing concrete realizations of 
the mapping discussed in this work. What amounts to the same 
thing, they are worked out examples where the Caldeira-Legget model is 
explicitly derived. 

In this respect, in fact, it is important to note that any application of our 
mapping is concretely substantiated in the Caldeira-Leggett model. At 
the same time, we stress again that the present work is  
focused on the question of deeply understanding what is really behind 
the fact that this modelization is so successful in covering the essential 
features of dissipative systems. We believe that we achieve this goal by 
performing a thorough analysis of the fluctuation-dissipation theorem. 

In summary, we have found that when linear 
response theory applies, any generic system can be mapped onto a 
fictitious system of harmonic oscillators so that that the mean square 
$\langle \hat{A}^2(\omega)\rangle$ of the fluctuating observable and the 
corresponding imaginary part of the generalized susceptibility  
$\chi_{_A}^{''} (\omega)$ of the real system are given by the corresponding 
quantities of the fictitious one. 
Moreover, we have seen that such a mapping allows to consider the 
BE distribution factor which appears in the FDT as originating from the 
individual harmonic oscillators of the fictitious equivalent system. 
This strongly suggests that it is only in this sense that this BE 
factor can be interpreted in terms of harmonic oscillators and that no 
other physical meaning can be superimposed to it. 

We believe that our mapping has a broader range of applicability 
than the worked case of the FDT discussed in this paper. Work is in 
progress in this direction.  

\vskip 12pt

{\bf{\Large Acknowledgements}}
\vskip 12pt

We thank Luigi Amico, Marcello Baldo, Pino Falci, J.C. Taylor and 
Dario Zappal\`a for many useful discussions.

\end{document}